\documentclass[
showpacs,
 amsmath,amssymb,
 aps,
 10pt,
twocolumn
]{revtex4-1}

\usepackage[usenames,dvipsnames]{color}
\usepackage{graphicx,textcase} % Include figure files
\usepackage{dcolumn,overpic} % Align table columns on decimal point
\usepackage{bm,color}% bold math
\usepackage[T1]{fontenc}%needed in order to use Polish Characters !!!
\usepackage{ae,aecompl}%to make it all look better using T1
\usepackage{chngcntr}
\usepackage[para]{threeparttable}
\usepackage{etoolbox}
\usepackage{lipsum}
\usepackage[bottom]{footmisc}
\usepackage{setspace}
\newcommand\T{\rule{0pt}{2.6ex}}       % Top strut
\AtEndEnvironment{table*}{\vskip10pt}{}{}% change here

\usepackage{url}
\usepackage{hyperref}
\hypersetup{colorlinks=true,breaklinks,linkcolor=blue,urlcolor=blue,citecolor=blue}

\begin{document}

\title{Electronic structure, magnetism and exchange integrals in transition metal oxides: role of the spin polarization of the functional in DFT+$U$ calculations}
\author{Samara~Keshavarz$^1$, Johan~Sch\"ott$^1$, Andrew J. Millis$^{2,3}$, and Yaroslav~O.~Kvashnin$^1$}

\affiliation{$^1$Uppsala University, Department of Physics and Astronomy, Division of Materials Theory, Box 516, SE-751 20 Uppsala, Sweden}
\affiliation{$^2$Department of Physics, Columbia University, New York, New York, 10027, USA}
\affiliation{$^3$Center for Computational Quantum Physics, The Flatiron Institute, New York, NY 10010}
\date{\today}

\begin{abstract}
Density functional theory augmented with Hubbard-$U$ corrections (DFT+$U$) is currently one of the widely used methods for first-principles electronic structure modeling of insulating transition metal oxides (TMOs).
Since $U$ is relatively large compared to band widths, the magnetic excitations in TMOs are expected to be well described by a Heisenberg model. However, in practice the calculated exchange parameters $J_{ij}$ depend on the magnetic configuration from which they are extracted and on the functional used to compute them. In this work we investigate how the spin polarization dependence of the underlying exchange-correlation functional influences the calculated magnetic exchange constants of TMOs. We  perform a systematic study of the predictions  of calculations based on the  local density approximation plus $U$ (LDA+$U$) and the local spin density approximation plus $U$ (LSDA+$U$) for the electronic structures, total energies and magnetic exchange interactions $J_{ij}$'s extracted from ferromagnetic (FM) and antiferromagnetic (AFM) configurations of several transition metal oxide materials. We report that, for realistic choices of Hubbard $U$ and Hund's $J$ parameters, LSDA+$U$ and LDA+$U$ calculations result in different values of the magnetic exchange constants and band gap. The dependence of the band gap on the magnetic configuration is stronger in LDA+$U$ than in LSDA+$U$ and we argue that this is the main reason why the configuration dependence of the  $J_{ij}$'s is found to be systematically more pronounced in LDA+$U$ than in LSDA+$U$ calculations. We report a very good correspondence between the computed total energies and the parameterized Heisenberg model for LDA+$U$ calculations, but not for LSDA+$U$, suggesting that LDA+$U$ is a more appropriate method for estimating exchange interactions. 
\end{abstract}

\maketitle

\section{Introduction\label{introduction}}
Transition metal oxides (TMOs) possess a plethora of interesting physical properties. A classical example is the outstandingly rich phase diagram of doped manganites~\cite{cmo-phasediag-orig, cmo-phasediag}. First-principles modelling of transition metal oxides and related materials is of longstanding interest, and their potential as multiferroics and for antiferromagnetic spintronics ~\cite{afm-spintronics} as well as the increasing importance of high throughput calculations (for example in materials genome-related studies)  creates an urgent need to understand the strengths and weaknesses of different computational methods in providing descriptions of the main properties of TMOs.

The complex correlated nature of the TMO electronic ground states have led to the development and extension of electronic structure methods~\cite{tmos-hybrid-2004,mno-hybrid,PhysRevB.85.054417,C6RA21465G,Park15,lda-lsda}. Density functional theory augmented with a Hubbard $U$ correction~\cite{ldau1,lsdau,ldau2} (DFT+$U$) is currently the most widely used first-principles method for studying TMOs because of its clear physical formulation and limited computational costs, which allow to tackle even large systems. Within this theory, one selects certain \textit{correlated} orbitals, and explicitly considers local Coulomb interactions for them.  Alternatively, self-interaction-corrected DFT~\cite{sic-tmos} and hybrid functionals~\cite{hybrid-fncls,hybrid-functionals} are also commonly used~\cite{sic-tmos2,PhysRevB.85.054417,mno-hybrid} and often provide a physical picture rather similar to that of DFT+$U$. More sophisticated approaches based on GW approximation~\cite{hedin-GW,gw-ldau}, dynamical mean-field theory~\cite{kotliar-DMFT} or a combination of the two~\cite{gw-dmft-review} are employed in TMOs studies, but are more computationally demanding and their development is currently ongoing.

A basic question  is whether to base augmented density functional theory on an exchange correlation functional that depends on charge density only \cite{ldau1}  or on an exchange-correlation functional that depends on both charge and spin density ~\cite{lsdau}. In the rest of the paper we will refer to these two methods as LDA+$U$ and LSDA+$U$ functionals, respectively\footnote{The notation of DFT+$U$ and sDFT+$U$ would be more appropriate here, since we believe the conclusions of this work would also hold for the functionals based on generalized gradient approximation (GGA). However, the  notation used here is more common in the literature  and is less misleading}. In LDA+$U$, the direct energetic contribution to the magnetism originates entirely from the Hubbard $U$ term and thus appears exclusively on the subset of orbitals considered as \textit{correlated}. The remaining orbitals may acquire non-zero inducted magnetism from hybridization with the correlated orbitals. On the other hand, in the LSDA+$U$ approximation the spin polarization of all orbitals contributes directly to the magnetic energy. This is widely believed to be a more reasonable approximation, and for this reason calculations based on LSDA+$U$ are more common. 

Recently, Chen and Millis~\cite{lda-lsda} building on previous work of Marianetti, Park and Millis ~\cite{Park15} compared  the total energies, magnetic moments and density of states  for several TMO materials predicted by  spin-dependent and spin-independent functionals. The results were also analyzed in terms of strength of $U$ and compared on a qualitative level with the Slater-Kanamori Hamiltonian. One of the main conclusion of this study was that the onsite exchange splitting of the transition metal $d$-orbitals  is overestimated in LSDA+$U$. 

Many TMOs have  magnetic ground states; in insulating TMOs the low energy magnetic excitations are expected to be described by a Heisenberg-type model and theoretical estimates of the values of the exchange couplings in these models are of interest. The exchange couplings in TMOs have been addressed on a first-principles level by numerous studies~\cite{NiO-jij-SIC, TMO-Halle, Jacobsson-TMOs, Logemann-2017,u-tmo2,mno-hybrid,j-tmo-2,j-tmo-3}. They may be obtained from the electronic structure methods by computing total energy differences  between different magnetic states,  by use of the magnetic force theorem (MFT)~\cite{jij-1} or from spin spiral calculations~\cite{feconi-halilov}. These three methods provide different results (see e.g. Refs.~\cite{TMO-Halle, Jacobsson-TMOs, Logemann-2017,lda-lsda-korea-2}). Moreover, the MFT-derived $J_{ij}$'s depend on the magnetic ground state from which they are extracted. It has been argued ~\cite{Logemann-2017} that correlated oxides with relatively large $U$ values can be expected to be prototypical Heisenberg magnets characterized by configuration-independent exchange interaction although at large $U$ charge transfer processes involving the ligand ions and controlled by the band gap become more important \cite{Zaanen85} so ratio of the bandwidth to the charge transfer gap may be a more relevant criterion. In several works, the configuration-dependence of the MFT-derived $J_{ij}$ is argued to be a signature of higher-order magnetic interactions~\cite{antropov-nat11, yk-fepd3}, in line with the work of Fedorova \textit{et al}~\cite{fedorova-biquad}, highlighting the importance of biquadratic interactions in manganites. Another reason for the configuration dependence of the interactions in TMOs is argued to be the oxygen polarization, which is present in the FM state, but is absent in the AFM one~\cite{solovyev-review-magnanites,CrO2-O-polar}. Logemann and co-workers~\cite{Logemann-2017} found that explicit inclusion of  the oxygen magnetic moments reduces the state dependence of the calculated exchange constants for NiO, but not for MnO, indicating that explicit inclusion of ligand polarization does not systematically improve the description of magnetic properties of TMOs.

In summary, the previous literature indicates that the calculated exchange constants of transition metal oxides depend on the methodology used for the calculation and on the magnetic state, calling into question an interpretation of the magnetic properties in terms of a Heisenberg-like model. This paper characterizes the  differences between results of LDA+$U$ and LSDA+$U$ calculations for the energies, band gaps, and exchange constants of the transition metal oxide compounds  CaMnO$_3$ (CMO), MnO, FeO, CoO and NiO. We have chosen these systems, which have different crystal structures and different $d$-level occupancies,  to ensure the generality of our conclusions. Our results enable the  identification of  the origin of the observed deviations from Heisenberg model of the TMOs and indicate that functionals of density only (rather than spin density dependent functionals) provides a  reasonable description of the magnetic energetics, consistent with a Heisenberg-like description. 

The article is organized as follows.  In Section~\ref{theory}, we provide the theoretical background and the details of the computations performed in this work. The main results are presented in Section~\ref{result}, which compares calculated properties obtained with two different functionals. The results are discussed, and conclusions drawn,  in  Section~\ref{conclusions}.

\section{Computational Details}
\label{theory}

We study CaMnO$_3$ (CMO), MnO, FeO, CoO and NiO. Relevant parameters including space groups and lattice constants are shown in  Table~\ref{tab:uandj}. While MnO and NiO are rhombohedral below their N\'eel temperatures, FeO and CoO exhibit larger-amplitude distortions from the rock-salt structure due to an orbital-ordering induced by the Jahn-Teller effect and leading to a further reduction of the crystal symmetry to monoclinic. 
However, the change in the lattice parameters is not substantial, keeping them to be less than 1\% different. All of the compounds order antiferromagnetically at low temperatures. For the monoxides the antiferromagnetism is of the N\'eel type characterized by opposite spins on adjacent TM sites, this may also be viewed as a stacking of ferromagnetic planes along (111) axis with alternating direction of the TM moments~\cite{feo1}. CMO is different and has an orthorhombic structure with $b\simeq 0.99a$ and large $c$, possessing a $G$-type AFM order.

We have computed the electronic structures of all of the compounds using  the full-potential linear muffin-tin orbital (FP-LMTO) code RSPt~\cite{rspt-web};  implementation details at the DFT level are described in Ref.~\cite{rspt-book} and the implementation of the DFT+$U$ formalism is given in Refs.~\cite{rspt-lda+u,rspt-oscar}. According to this formalism, the energy functional is given by:
\begin{equation}
\label{ldau}
E_{\mathrm{DFT}+U}=E_{\mathrm{DFT}}+E_U-E_{\mathrm{DC}}
\end{equation}
where $E_{\mathrm{DFT}}$ is the contribution to the total energy from one the exchange-correlation functionals, e.g., local density approximation (LDA) in our case.  $E_U$ is the energy contribution due to the Hubbard term, added in a static mean-field fashion. Finally, $E_{\mathrm{DC}}$ is a compensation to avoid double-counting (DC) of the interactions contained both in $E_{U}$ and $E_{\mathrm{DFT}}$. 

The fully rotationally invariant formulation of the Hubbard interaction (4-index $U$-matrix~\cite{ldau2,4-index-2}) is used throughout this paper unless otherwise stated. This matrix can be parametrized via Slater integrals that are constructed from two effective parameters: onsite Coulomb $U$ and Hund's exchange $J$ as described in Ref.~\cite{ldau2}. These parameters can be either extracted from experiments or calculated from first-principles. In this work, we have taken their values from the literature as listed in Table.~\ref{tab:uandj}. For the case of CMO, the value of effective $U$ ($U_\mathrm{eff}=U-J$) is reported around 3 to 4~eV~\cite{u-cmo}. However, our previous study using $U=4$~eV and $J=0.9$~eV resulted in a magnetic moment and an ordering temperature in agreement with experiment~\cite{cmo-u}. For the other TMO system, the $U$ value can change from 6 to 9 eV depending on, e.g., the $d$ shell occupations~\cite{u-tmo1,u-tmo2}. The value of the Hund's exchange $J$, on the other hand, is typically around 0.9 eV and only weakly depends on the environment of the $3d$ atom in solids. Therefore, we set this value equal to 0.9 eV for all the TMO systems studied in this work.

We used the fully localized limit (FLL) formulation of the double counting, with the double counting potential~\cite{dc-fll} 
\begin{equation}
\label{dc}
v_{\mathrm{DC},\sigma} \equiv \frac{\partial E_\mathrm{DC}}{\partial n_\sigma} = U (n-\frac{1}{2}) - J (n_\sigma -\frac{1}{2})   
\end{equation}
where $n=n_{\uparrow}+n_{\downarrow}$ is the total number of electrons on the correlated orbitals (TM-$3d$ in the cases studied here).
To compute $E_{\mathrm{DFT}}$ we  used two exchange-correlation functionals: the local density approximation (LDA) and the local spin density approximation (LSDA). In the former case the DFT energy has no explicit dependence on the magnetization (of course there is an implicit dependence arising from band rearrangement in the magnetic state) while in the latter case the DFT energy depends explicitly on the local magnetization. This difference affects the double-counting correction. 
In LDA+$U$ the spin density does not enter the DFT functional correction, so that $n/2$ is used instead of $n_{\sigma}$ in Eq.~\eqref{dc}, and thus the correction is the same for both spin channels. 
In LSDA+$U$, the DC correction is different for states with opposite spin projections $\sigma$. 

The convergence of the results with respect to the number of $k$-points has been analyzed by performing selected calculations with large meshes. 
Finally, a converged $k$-mesh of 16$\times$16$\times$16 for MnO, FeO, Co and NiO, a mesh of 9$\times$9$\times$7 for CMO have been employed.

\begin{table}
\caption{\label{tab:uandj} The space group (SG), the experimental lattice parameters (\AA) as well as the Hubbard $U$ (eV) and the Hund's exchange $J$ (eV) parameters used for each compound considered in this work.}
\begin{ruledtabular}
\begin{tabular}{lccccccc} 
  & CMO & MnO & FeO &  CoO  &   NiO   \T  \\
\hline 
SG   & $Pnma$ & $R\bar{3}m$ & $C2/m$ & $C2/m$ & $R\bar{3}m$  \T  \\
\hline
$a$ & 5.29\footnotemark[1] & 4.45\footnotemark[2] & 4.32\footnotemark[3] & 4.26\footnotemark[2] & 4.17\footnotemark[2]  \T \\
$b$ & 5.28 & &  & &   \\
$c$ & 7.46 & & &  &  \\
\hline
$U$ & 4.0 & 6.0 & 7.0 & 7.0 & 8.0   \T  \\
$J$  & 0.9 & 0.9 & 0.9 & 0.9 & 0.9  \\
\end{tabular}
\begin{tablenotes}
\item[a] Reference~\cite{l-cmo}. 
\item[b] Reference.~\cite{l-mno}. 
\item[c] Reference.~\cite{l-feo}. 
 \end{tablenotes} 
\end{ruledtabular}
\end{table}

The magnetic force theorem~\cite{jij-1,jij-2000-lk} is applied to the converged DFT+$U$ solutions, leading to estimates for the pairwise exchange constants $J_{ij}$ between the two spins, located at sites $i$ and $j$ and providing a mapping of the magnetic excitations onto the Heisenberg Hamiltonian: 
\begin{equation}
\hat{H}=-\sum_{i\neq j}J_{ij}\vec{e}_i\cdot\vec{e}_j,
\label{equ:heisen}
\end{equation}
$\vec{e}_i$ is a unit vector along the magnetisation direction at  site $i$. Negative (positive) sign of the $J_{ij}$ corresponds to AFM (FM) coupling. More details of the evaluation of the exchange interactions, particularly in relation to the basis set used for the local orbitals, can be found in Ref.~\cite{jij-2}.

\section{results and discussions}
\label{result}

The transition metal oxide compounds listed in Table~\ref{tab:uandj} have different electronic structures and local physics. In all cases the local environment leads to an approximately cubic point symmetry for the transition metal ion. In CaMnO$_3$ the transition metal formal valence is high-spin $d^3$; the $t_{2g}$ shell is half filled and $e_g$ shell is empty and the insulating gap separates the (majority spin) $t_{2g}$ and $e_g$ manifolds. In MnO the transition metal formal valence is high-spin $d^5$, with half filled $t_{2g}$ and $e_g$ shells and the insulating gap separates the majority and minority spin manifolds. In NiO (high-spin $d^8$) the $t_{2g}$ shell is fully filled, the $e_g$ shell is half filled, and the gap separates majority and minority spin $e_g$ states. 
The situation is different in the case of FeO and CoO. In these compounds  the $t_{2g}$ shell is partially occupied and orbital order may occur in the ground state. 
The theoretically obtained gaps, spin and orbital moments are also different in the literature~\cite{mazin,kamal,schron,solov}, which indicate the emergence of the local minima in system. In order to avoid converging to a local minima rather than the ground state, we have done a systematic study of the local minima in these systems by means of occupation matrix control. 
These results will be published elsewhere. Here we only present results for the global ground states for fixed magnetic structure.

\begin{figure}[t]
\includegraphics[width=\columnwidth]{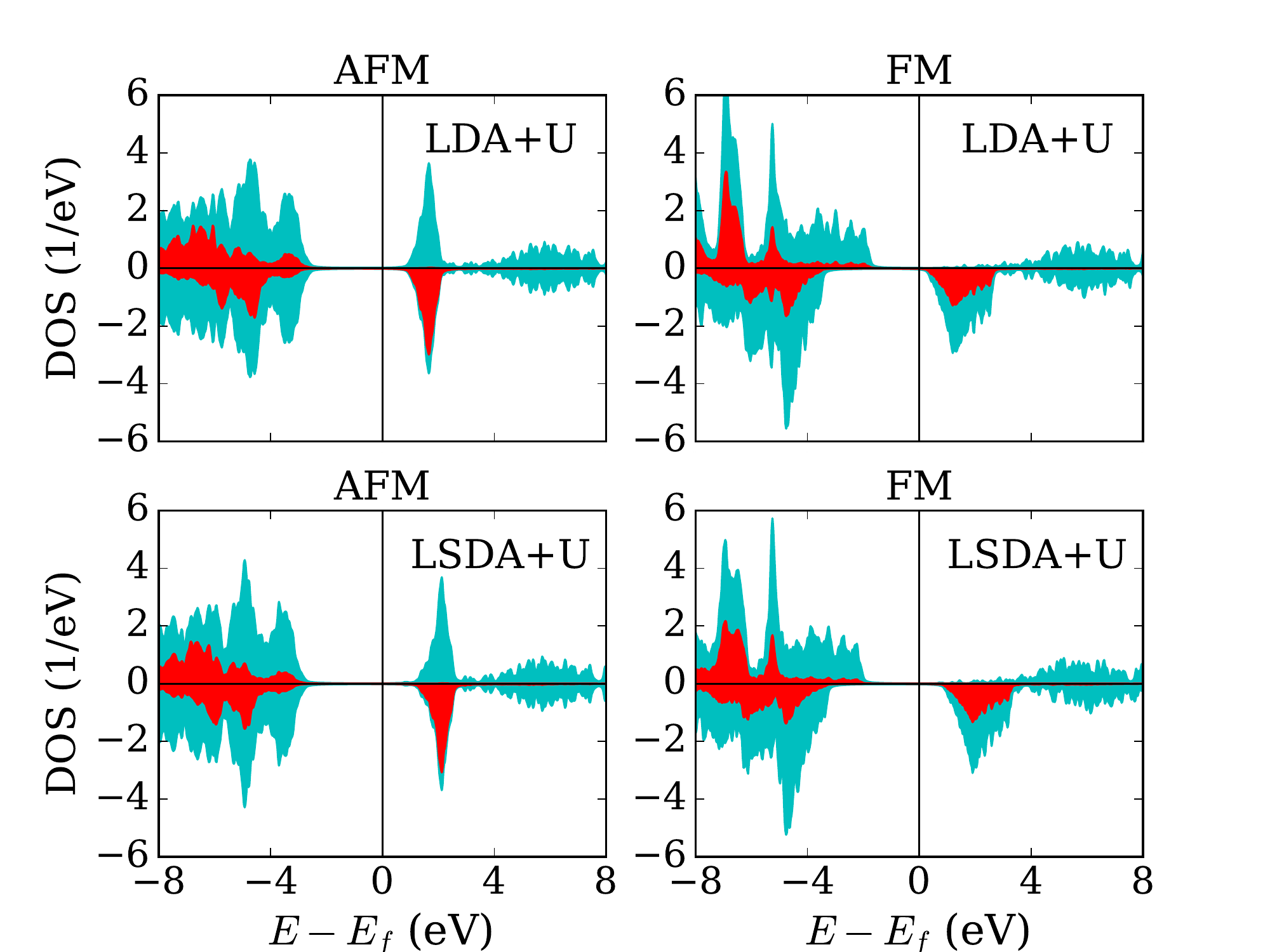}
\caption{Total density of states (cyan) as well as the density of states projected onto the Ni 3$d$ orbitals (red). }
\label{fig:dos-nio}
\end{figure}

We begin with NiO. Fig.~\ref{fig:dos-nio} presents the total density of states as well as the projected density of states of the $d$ orbitals in LDA+$U$ and in LSDA+$U$ methods for AFM and FM states. The system is found to be insulating for both states and both methods. The AFM state is the ground state of the system with the total energy of 0.26 eV (0.16 eV) lower than that of the FM state in LDA+$U$ (LSDA+$U$). The gap in the FM state is slightly smaller than the gap in the AFM state in both methods. This is expected as also has been discussed in Ref.~\cite{dalton} for a model system.

The first and second (neighbor exchange interactions found via the MFT method are shown in Tab.~\ref{tab:jij-nio}. The  dominant interaction is the second neighbor exchange $J_2$. The first neighbor interaction is extremely small because there is no superexchange path between the $e_g$ orbitals through one anion in this geometry (see e.g. Ref.~\cite{keffer}). We see that the exchange interactions are negative, favoring  antiferromagnetism as found also in the DFT total energy calculations. However the values of the exchange constants  depend on the state and computational method, being strongest when computed in the ferromagnetic LDA+$U$ ground state and weakest when computed from the antiferromagnetic LSDA+$U$ state. Comparing Fig.~\ref{fig:dos-nio} and Tab.~\ref{tab:jij-nio} reveals that a larger band gap is associated with weaker exchange interactions, as expected for super-exchange interaction in insulating TMOs. In  Eq.~\eqref{equ:heisen}  the values of the onsite magnetic moments are effectively encoded in the $J_{ij}$'s, one may ask whether the observed differences in exchange constants arise from a difference in magnetic moments. The results presented in the right-most column of Table.~\ref{tab:spinmom} show that this is not the case.  

\begin{table}[t]
\caption{\label{tab:jij-nio} Exchange parameters (meV) in NiO, for FM or AFM reference states using LDA+$U$ or LSDA+$U$ methods. R refers to the reference state, whereas M refers to the method.}
\begin{ruledtabular}
\begin{tabular}{lcccc} 
R & AFM        & AFM        & FM        &  FM  \\
M & LDA+$U$ & LSDA+$U$ & LDA+$U$ &  LSDA+$U$   \T  \\
\hline 
$J_1$   & \: 0.01 & -0.03  &   \:\:\:  0.01  & \: -0.05    \T  \\
$J_2$   & -9.27  & -7.46  & -12.54    & -8.79      \\
\end{tabular}
\end{ruledtabular}
\end{table}

\begin{figure}[t]  
\includegraphics[width=\columnwidth]{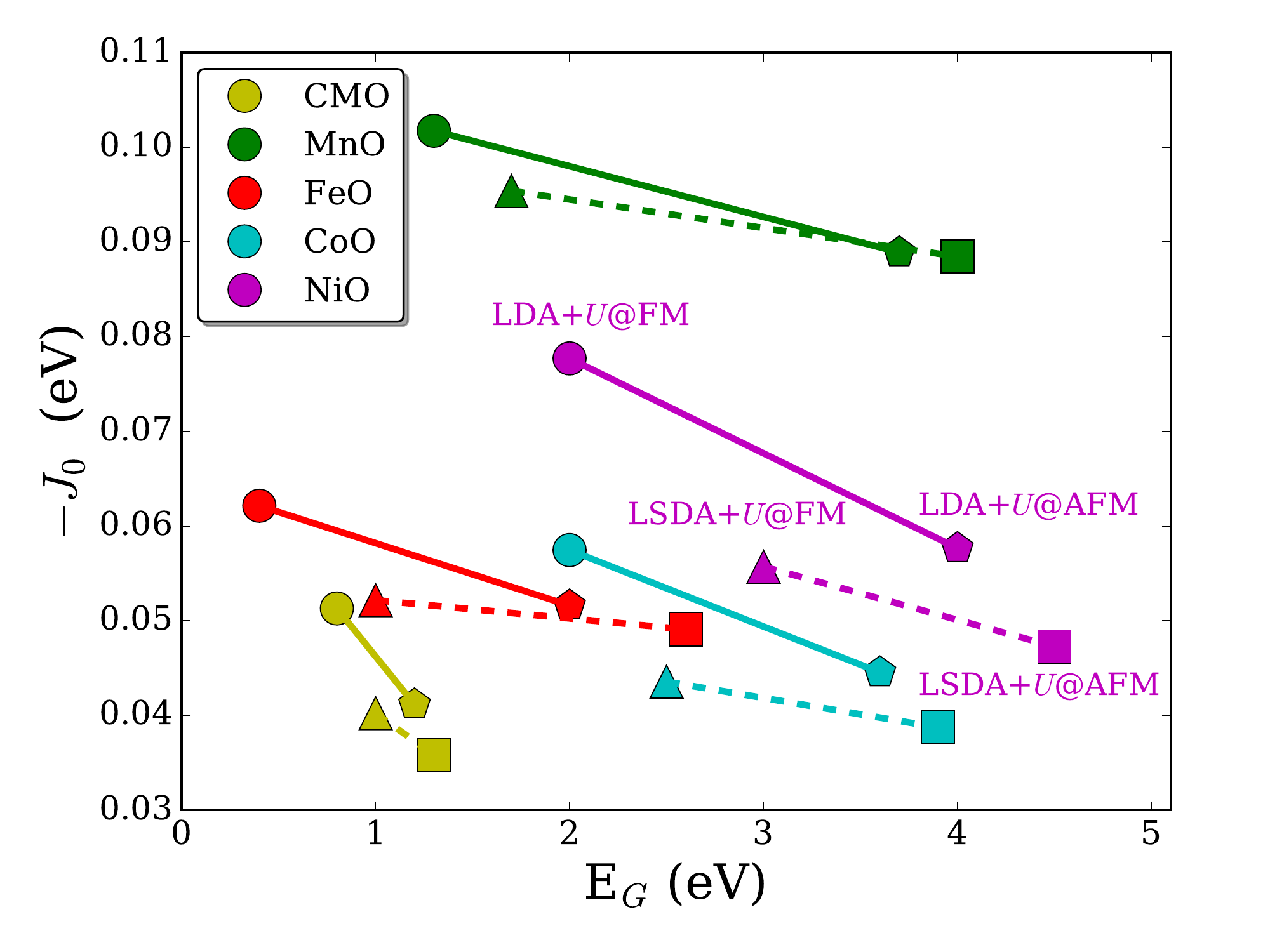}
\caption{Total exchange field ($J_0$) plotted against calculated band gap ($E_G$) calculated using DFT methods and reference states shown. 
Because the net exchange is antiferromagnetic we plot $-J_0$ for ease of reading. 
Circles (pentagons) represent the LDA+$U$ results based on FM (AFM) state and triangle (square) represent LSDA+$U$. 
Solid lines connect LDA+$U$ results obtained in FM states to those obtained in AFM states and dashed lines connect the FM and AFM LSDA+$U$ results.
The experimental band gaps are 1.55 eV~\cite{Loshkareva2012}, 3.9 eV~\cite{PhysRevB.44.1530}, 2.4 eV~\cite{BOWEN1975355}, 2.5 eV~\cite{PhysRevB.44.6090} and 4.3 eV~\cite{PhysRevLett.53.2339} for CMO, MnO, FeO, CoO and NiO, respectively.
}
\label{fig:j-gap}
\end{figure}

We now compare results across systems. Figure~\ref{fig:j-gap} plots the total exchange field (summed over all neighbors) $J_0=\sum_{j}J_{0j}$ against the band gap for all the systems studied here, using both methods and both reference states.  We see that the band gaps are larger, and exchange constants smaller, for the AFM state than for the FM state, and band gaps are larger and exchange constants smaller in LSDA+U than in LDA+U.  
The reference dependence of band gap and exchange constants is stronger than the dependence on the method, and is more pronounced in LDA$+U$ than in LSDA$+U$.
Comparison of the slopes of the connecting lines emphasizes that the $J_0$ is more sensitive to the size of the gap in LDA+$U$ (solid lines) than in LSDA+$U$ (dashed lines) approach. 
However, in no case is the dependence of $J_0$ on $E_G$  as strong as the $J_0\propto E_G^{-1}$ expected from simple superexchange arguments. 

\begin{table}
\caption{\label{tab:spinmom} Spin moment ($\mu_B$) of the 3$d$ atom as well as the oxygen atom in each system. R refers to the reference state (FM or AFM ground state), whereas M refers to the method (LDA+$U$ or LSDA+$U$).}
\begin{ruledtabular}
\begin{tabular}{llccccccc} 
&  & CMO & MnO & FeO &  CoO  &   NiO   \T  \\
\hline 
R & \:\:\:\: M & & 3$d$  & &  \T  \\
\hline
AFM & LDA+$U$    & 2.54  & 4.55  &   3.64  &  2.70  &    1.72  \T   \\
AFM & LSDA+$U$  & 2.60  & 4.56  &   3.64  &  2.73  &  1.76    \\
FM   & LDA+$U$      & 2.64  & 4.59  &   3.69  &  2.74  &    1.77   \\
FM   & LSDA+$U$    & 2.65  & 4.58  &   3.66  &  2.75  &    1.79     \\
\hline 
 & & & O & &  \T  \\
\hline
FM & LDA+$U$      &   0.06    &  0.19  &  0.18    &  0.16   &   0.16  \T   \\
FM & LSDA+$U$    &   0.05    &  0.12  &  0.13    &  0.13   &   0.14      \\
\end{tabular}
\end{ruledtabular}
\end{table}

\begin{figure}[t]
\includegraphics[width=\columnwidth]{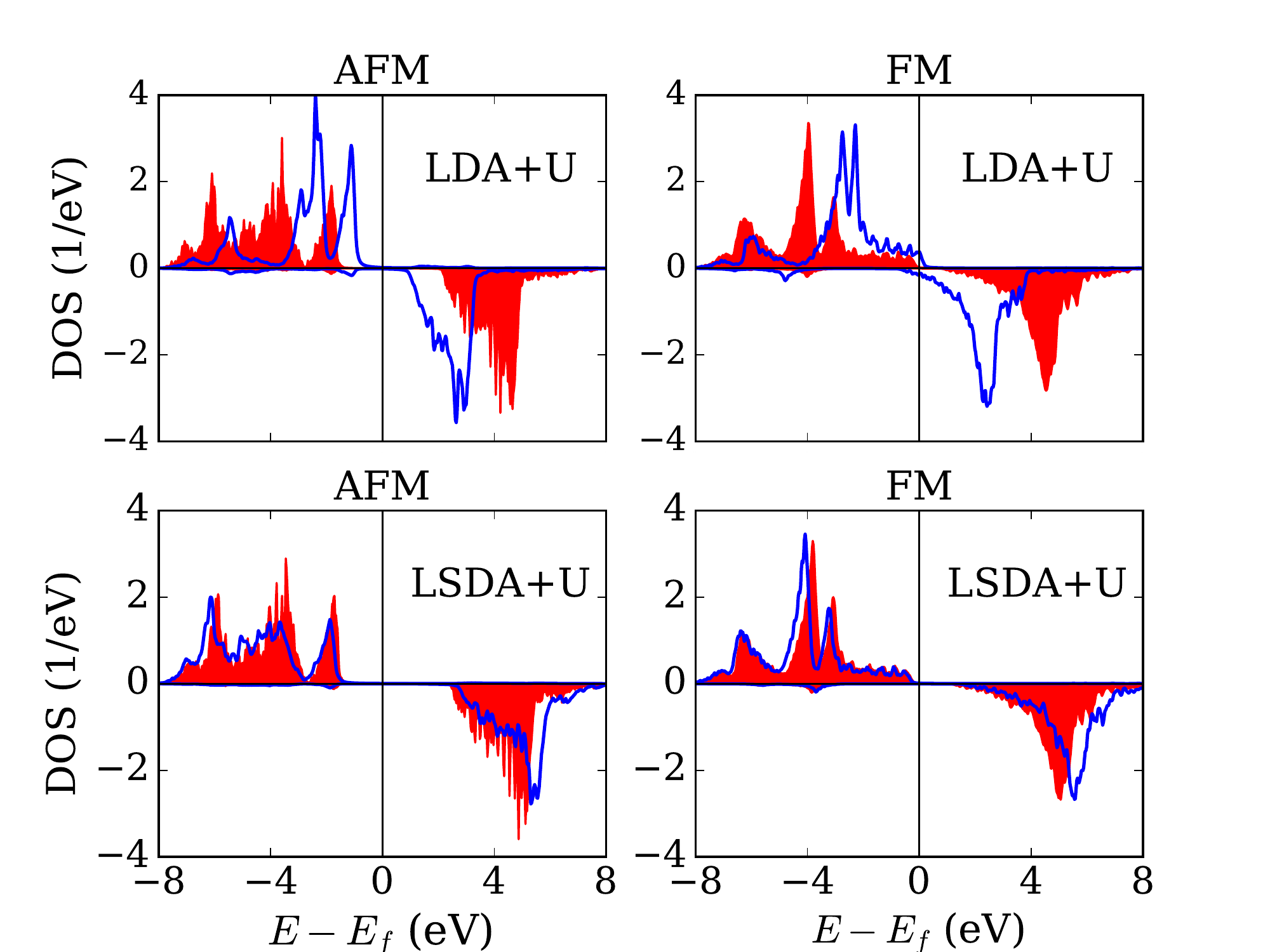}
\caption{Projected density of states of 3$d$ orbitals in MnO for $U=6$ eV, $J=0.9$ eV (red) and for $U=6$ and $J=0.0$ eV (blue lines).}
\label{fig:pdos-mno}
\end{figure}

In the LDA+$U$ calculations reported here, magnetism arises from the additional interactions  $U$ and  $J$, which favor high spin configurations of the transition metal $d$ orbitals. In the LSDA+$U$ calculations there are additional contributions to magnetism from the spin dependence of the exchange correlation functional and the double counting ($J$ and $m$-dependent in LSDA+$U$ but not LDA+$U$). To understand these additional effects, we have performed a set of calculations in which the Hund's exchange $J$ is set to 0. In this way, the double counting term is the same for both LDA+$U$ and LSDA+$U$ methods so that  difference in the results stems from the XC functional. Fig.~\ref{fig:pdos-mno} shows  one example: the projected density of states for the $d$-orbitals for MnO, (half filled $d$-shell). We see that for MnO in both AFM and FM states, setting $J=0$ in the LDA+$U$ case reduces the splitting between majority and minority $d$-states and also the computed gap (indeed in the ferromagnetic case the majority and minority spin bands overlap slightly, leading to a metallic state), whereas for LSDA+$U$ setting $J=0$ leads to a slight increase in exchange splitting and gap size. 
The LDA+$U$ result, that adding a $J$ increases the exchange splitting of the transition metal $d$-states, seems intuitively reasonable~\cite{lda-lsda}. 
However, nonzero $J$ often tends to decrease the exchange splitting in LSDA+$U$~\cite{lda-lsda}, but it might not be true in general~\cite{lda-lsda-korea-1}.

\begin{figure}[b]
\includegraphics[width=\columnwidth]{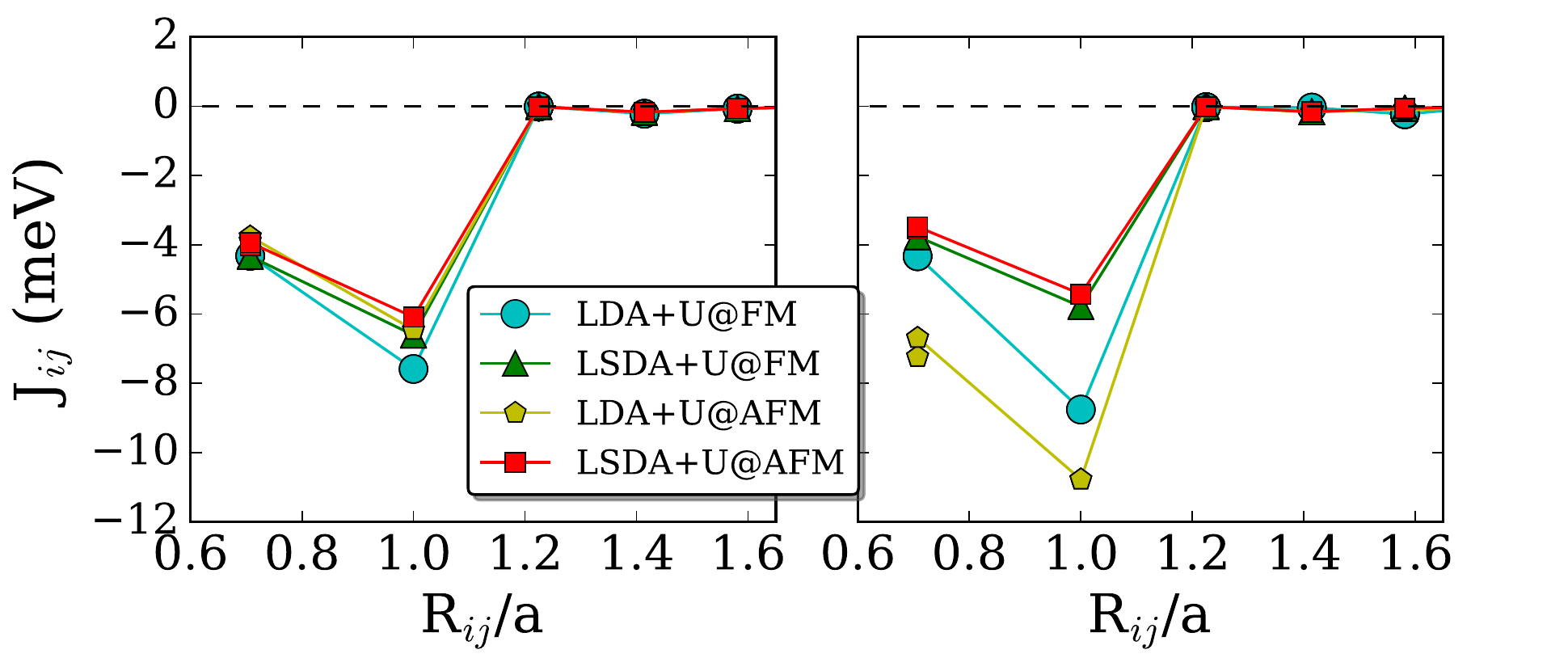}
\caption{Exchange parameters  computed for MnO using LDA+$U$ and LSDA+$U$ methodology for $U=6$ eV and $J=0.9$ eV (left panel) and $J=0.0$ eV (right panel) plotted against interionic distances.}
\label{fig:J-mno}
\end{figure}

The corresponding exchange parameters computed for MnO with and without $J$ are shown in Fig.~\ref{fig:J-mno}. Comparison of the left ($J=0.9$ eV) and right ($J=0$) panels shows that in the LDA+$U$ approximation decreasing the onsite $J$ increases the magnitude of the exchange constants (especially the crucial second neighbor constant $J_2$), as expected from the decrease of the insulating gap. However, in the LSDA+$U$ approximation setting $J=0$ leads to a slight decrease in the magnitude of the exchange constants, consistent again with the slight increase in the gap.  Interestingly, in the LDA+$U$ case the increase in the magntiude of the intersite exchange on setting $J=0$ is noticeably  greater for the  AFM state than for the FM state.  The weaker dependence of the intersite exchange on $J$ in the LDA ferromagnetic state arises because at $J=0$ the ground state is metallic, so a superexchange picture is not valid.

\section{Conclusions}
\label{conclusions}

\begin{figure}[t]
\includegraphics[width=\columnwidth]{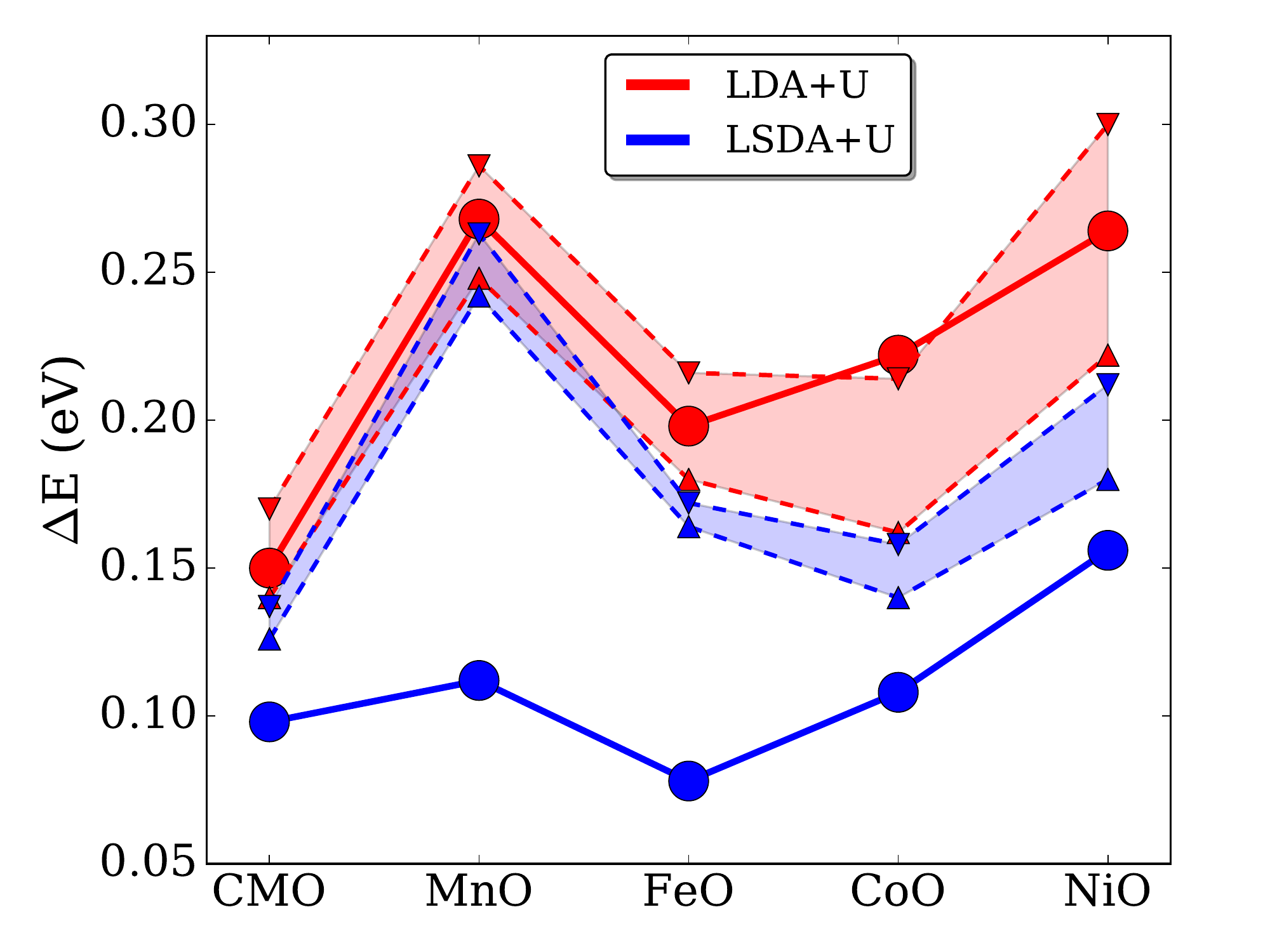}
\caption{Total energy differences ($\Delta E=E_{\textrm{FM}}-E_{\textrm{AFM}}$) obtained by LDA+$U$ (solid red line) and LSDA+$U$ (solid blue line) with those which are obtained via $J_{ij}$'s. The latter results are shown with dashed lines as follows: upward (downward) triangle for AFM (FM) in red for LDA+$U$ and blue for LSDA+$U$ methods. 
The colored areas are shown as the interval in which the total energy difference based on the $J_{ij}$'s of any reference state (FM, AFM or a combination of both) could end up.}
\label{fig:deltae}
\end{figure}

In order to gain insight into the ability of augmented density functionals to predict magnetic behavior we used the magnetic force theorem  to calculate intersite spin exchange interactions in a series of insulating transition metal monoxides and the perovskite CaMnO$_3$.  We considered both ferromagnetic and antiferromagnetic ground states and performed calculations based on the local density and local spin density approximations, augmented by onsite interactions coupling the transition metal $d$ orbitals. Because in the materials we consider, the charge gaps ($\sim$ eV) are large compared to the maximum spin wave energies ($\sim 0.2$ eV) it seems reasonable to expect the magnetic properties to be governed by a generalized Heisenberg model, with spin exchange terms that are independent of the reference state (ferromagnetic or antiferromagnetic) used to obtain them. We find, however, that the exchange interactions extracted from ferromagnetic reference states are larger than those extracted from antiferromagnetic reference states, although the dependence on reference state is noticeably weaker in the LSDA+$U$ approximation than in the LDA+$U$ approximation.  

Despite the reference-state dependence, a Heisenberg model description is not completely unreasonable. To demonstrate this, we consider the energy difference between FM and AFM states ($\Delta E$) in DFT+$U$ and compare them with the ones obtained using Eq.~\eqref{equ:heisen} with the calculated $J_{ij}$'s. For example, in case of CMO, where the AFM order is of $G$ type, the Heisenberg model yields the following total energies:
\begin{equation*}
\begin{aligned}
&E_{FM}=-6J_1-12J_2-8J_3-6J_4-... \\
&E_{AFM}=6J_1-12J_2+8J_3-6J_4+...
\end{aligned}
\label{equ:jij-energy}
\end{equation*}
where the $J_{ij}$'s are obtained through four different considered cases (combinations of methods and reference states). In Fig.~\ref{fig:deltae} we show the energy differences computed directly from LDA+$U$ and LSDA+$U$ total energies (circles connected by solid lines). In addition, the smaller triangles connected by dashed lines represent the energy differences computed through the intersite exchange constants as discussed above. 
The reference state dependence of the exchange constants implies uncertainty in the energy differences, which we represent as shaded areas bounded below by upward triangles (the AFM-derived exchange constants) and bounded above by downward triangles (the FM-derived exchange constants). Remarkably, we see that the directly computed LDA+$U$ energy difference falls within the range of uncertainty of the Heisenberg-based estimate, and has essentially the same trends with material, suggesting that the LDA+$U$ scheme indeed leads to the expected Heisenberg-like magnetic physics (albeit with the uncertainties in the precise values of the exchange interactions). 
On the other hand for the LSDA+$U$ approximation the directly-derived FM-AFM energy differences are both substantially smaller than the Heisenberg-derived values and do not show the same trends with material. 
In other words, the directly computed  LSDA+$U$ FM-AFM energy differences are not in correspondence with any set of extract of $J_{ij}$ parameters, implying that the LSDA+$U$  FM-AFM total energy difference contains additional contributions, which are not captured by the Eq.~\eqref{equ:heisen} with TM spin degrees of freedom.

\begin{figure}[t]
\includegraphics[width=\columnwidth]{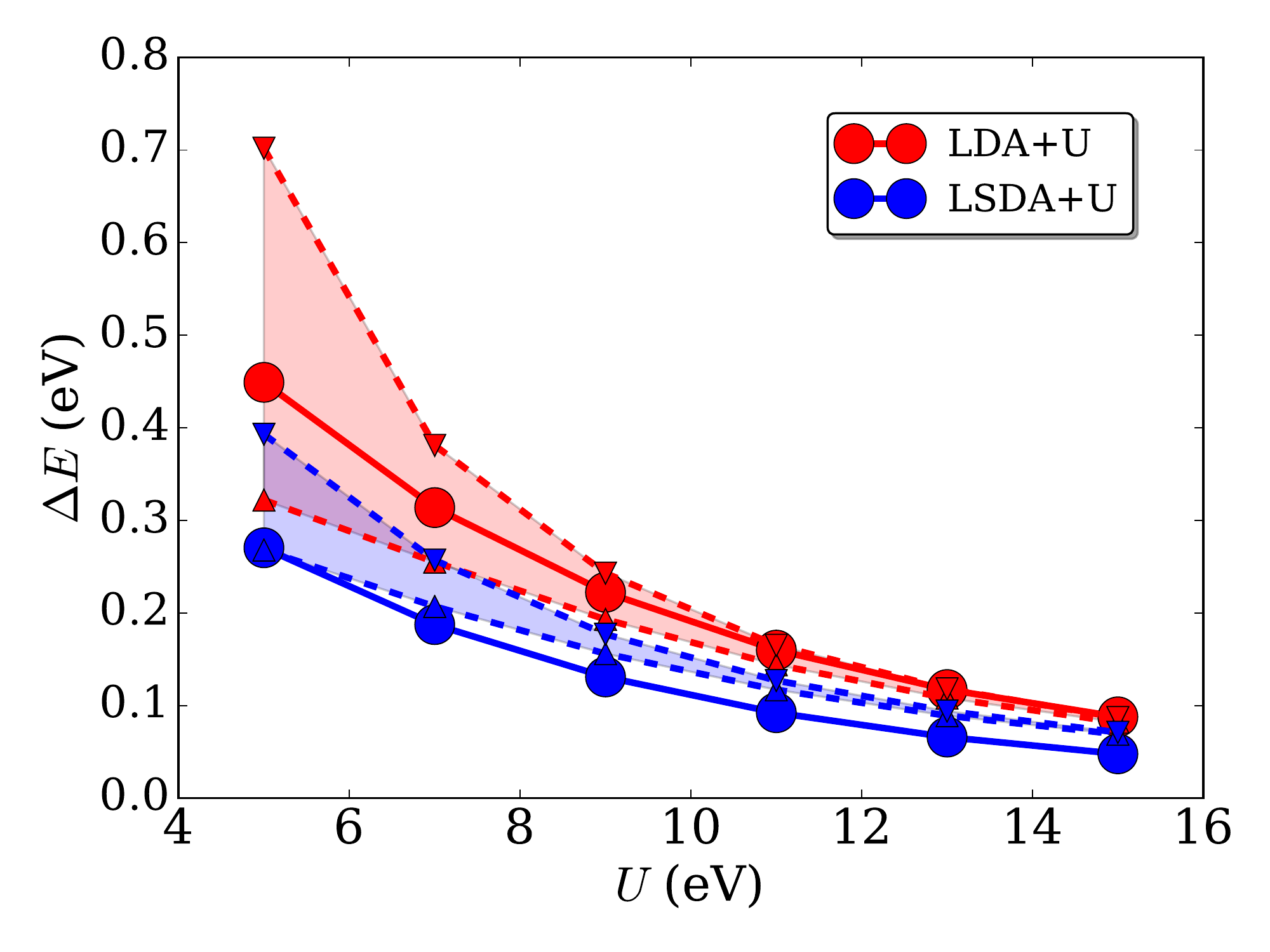}
\caption{Total energy differences ($\Delta E=E_{\textrm{FM}}-E_{\textrm{AFM}}$) obtained by LDA+$U$ (solid red line) and LSDA+$U$ (solid blue line) with those which are obtained via $J_{ij}$'s as a function of the $U$ value for the case of NiO. The latter results are shown with dashed lines as follows: upward (downward) triangle for AFM (FM) in red for LDA+$U$ and blue for LSDA+$U$ methods. 
The colored areas are shown as the interval in which the total energy difference based on the $J_{ij}$'s of any reference state (FM, AFM or a combination of both) could end up.}
\label{fig:deltae-2}
\end{figure}

The induced polarization on oxygen present in the ferromagnetic LSDA state might contribute to the discrepancy. However, the effect of oxygen spin polarization on the exchange and magnetic states has been studied for different systems~\cite{Logemann-2017,solovyev-review-magnanites,CrO2-O-polar}. This effect usually contributes to the FM energy as $\Delta E=E_{O-pol}+J_{ij}S_iS_j$ by means of the Stoner model as $E_{O-pol}=-IM^2/4$, where $I$ is the O spin splitting and $M$ is the O magnetic moment. Based on O moment (Tab.~\ref{tab:spinmom}), this term gives a value of 1 meV for CMO, 8 meV for MnO, 10 meV for FeO and CoO, 12 meV for NiO. These values are too small to account for the observed discrepancy in LSDA+$U$ results. An alternative source of the discrepancy might be the spin dependence of the  double-counting correction, which as shown in Fig.~\ref{fig:pdos-mno} and noted previously \cite{Chen15,lda-lsda}, leads to a counterintuitive dependence on the onsite Hund's coupling $J$ and (Fig.~\ref{fig:J-mno}) to a similarly counterintuitive  $J$-dependence of the LSDA-calculated exchange constants.

Finally, in Fig.~\ref{fig:deltae-2} we extend an analysis shown in Fig.~\ref{fig:deltae} by displaying the calculated $U$-dependence of the total energy differences in one of the systems, NiO.
The qualitative difference between LDA+$U$ and LSDA+$U$ become even more apparent with LDA+$U$ providing a consistent set between DFT- and Heisenberg model-based results.
It further supports and demonstrates the robustness of our conclusion regarding the spin-polarization of the functional.

In summary, the set of inter-site $J_{ij}$'s obtained by means of MFT provides good estimates of ordering temperatures and magnon dispersions, if combined with atomistic spin dynamics simulations~\cite{TMO-Halle,Jacobsson-TMOs,u-tmo2,Logemann-2017}. However, our results strongly suggest that if one extracts the $J_{ij}$'s from the DFT+$U$ total energies, the use of spin nonpolarized functional (e.g. LDA) is preferable. Moreover, in the light of this finding, it would be interesting to reconsider the importance of higher-order exchange interactions employing spin unpolarized DFT functional~\cite{fedorova-biquad}. 
We also demonstrate that the non-Heisenberg behaviour of the MFT-derived $J_{ij}$'s is a natural consequence of the different electronic structures of the FM and AFM states. We expect that the conclusions of the present study are not restricted to DFT+$U$ method and also hold for DFT+DMFT, which deserves a separate study. 

Note that alternatively one can extract exchange parameters in the paramagnetic (high temperature) phase, using the disordered local moments approach~\cite{dlm-1,dlm-2} or dynamical mean field theory~\cite{dmft-j}. These methods are complementary to those used here, which describe ground state and low-lying excitations. 

\section*{Acknowledgement}
We are thankful to Lars Nordstr\"{o}m and Patrik Thunstr\"{o}m for fruitful discussions.
S.K., J.S. and Y.O.K. acknowledge support from the Swedish Research Council (VR), eSSENCE and the KAW foundation. A.J.M acknowledges support from the Basic Energy Sciences Division of the U. S. Department of Energy via a subcontract from Argonne National Laboratories. The computer simulations are performed on computational resources provided by NSC, PDC and UPPMAX allocated by the Swedish National Infrastructure for Computing (SNIC). 

\bibliographystyle{apsrev4-1}

\end{document}